\documentclass{article}
\usepackage[utf8]{inputenc}
\usepackage{titling}
\usepackage{lipsum}
\usepackage[space]{grffile}
\usepackage{graphicx}
\usepackage{url}
\usepackage{amsmath}
\usepackage{amssymb}
\usepackage{multirow}

\usepackage{float}

\usepackage[scaled=1.1,p]{zlmtt}

\newcommand{\figwid}{0.99\linewidth}

\title{\bf Access-Adaptive Priority Search Tree}
\author{\Large {Haley Massa} and {Jeffrey Uhlmann} \\
Dept.\ of Electrical Engineering and Computer Science\\
University of Missouri - Columbia}
\date{}

\begin{document}

\maketitle

\begin{abstract}
In this paper we introduce the notion of explicit worst-case bounded adaptive algorithms for applications with fixed process-completion requirements. Such applications demand that a process be guaranteed to complete within an established time interval while adaptively reducing computational overhead during that interval, e.g., so as to reduce total energy usage. Our principal contribution is the {\em access-adaptive priority search tree} (AAPST), which can provide efficient distribution-sensitive performance comparable to the splay tree, but do so within strict -- and $O(\log n)$ optimal -- worst-case per-query bounds. More specifically, while the splay tree is conjectured to offer optimal adaptive amortized query complexity, it may require $O(n)$ for individual queries, whereas the AAPST offers competitive distribution-sensitive performance with strict $O(\log n)$ time complexity. This makes the AAPST more suitable for certain interactive (e.g., online and real-time) applications such as space system modules with reliability constraints involving rigid process-completion time intervals with secondary energy-minimization incentives.
\end{abstract}

\newpage

\section{Introduction}

Complex systems with strict reliability requirements (e.g., as defined by specified safety or performance constraints) may impose rigid completion windows on some or all processes. For example, a given process may be allotted $t$ units of time to complete, with failure to meet the deadline treated as a fault condition. Such a deadline is commonly defined to mitigate the impact of a divergent process on overall system performance by providing an unambiguous criterion for the signaling and subsequent handling of a fault condition. Of course, the time interval of the window for a given process must be confidently established so as to avoid spurious interrupts. The study of such systems became a priority in the early space program \cite{nasa} and are referred to as {\em time-critical tasks}:
\begin{quote}
\proptt{A time-critical task is characterized by repetitions of C units of processing over a frame interval F within a deadline time D $\leq$ F... The deadline, within which all processing must be completed, may result from a particular system implementation; that is, the output data must be available by D so that the system will have time to transfer the data out by the end of the frame. The deadline may also result from the requirements of the application itself.} \cite{tct}
\end{quote}
In many practical contexts, two different algorithms may be assessed as having equivalent utility if each can perform a particular operation within the same specified completion window, i.e., either algorithm is suitable for satisfying system requirements. However, in more stressing contexts in which energy consumption and/or heat generation must be minimized, e.g., for long-duration deep-space systems, there is need to optimize a more complex utility function. 
When evaluating prospective algorithms for such applications, there is typically a hard performance constraint that must be satisfied, plus a set of secondary performance attributes that are {\em preferred} but not required. We refer to the latter as {\em augmented complexity considerations}.

In this paper we present an efficient adaptive-access data structure and associated retrieval algorithm that can simultaneously provide worst-case optimal query complexity while also adaptively reducing computational costs -- which equates to minimizing energy consumption and heat generation -- compared to prior state-of-the-art alternatives under various operating assumptions. Specifically, we introduce the access-adaptive priority search tree (AAPST), which is based on the priority search tree of McCreight \cite{pst}, but with priorities representing key-access counts. We then demonstrate that our novel search algorithm offers strict optimal per-query performance bounds that cannot be supported using the state-of-the-art adaptive search tree, the splay tree \cite{splay}, while demanding less computational overhead on average than a balanced binary search tree (BST) in practical contexts in which the access distribution of keys is highly nonuniform.

\section{Background}

In most practical applications, the access of information, e.g., keys from a large dataset, is highly non-random. In other words, some keys will be queried much more frequently than others. This nonuniform sampling from a dataset of $n$ keys can potentially be exploited by a distribution-sensitive search structure to perform many fewer comparisons per query than required in the uniform case. For example, applications in which access frequencies are exponentially distributed, i.e., a small number of keys are accessed much more frequently than the remaining keys combined, an adaptive search algorithm and data structure may exhibit practical query performance that is much better than its $O(\log(n))$ complexity might suggest.

The splay tree \cite{splay} is a self-adjusting \cite{selfadj} binary search tree that optimizes its structure to the distribution patterns of the dataset. Splay trees differ from standard balanced BSTs by performing rotations that migrate frequently-accessed keys to the top of the tree so the search paths to those keys will be shorter when accessed during future queries. A novelty of the splay tree is that it does not necessarily enforce balance at all times during a given sequence of $n$ updates and/or queries, but it does guarantee that the complexity of any given sequence is $O(n\log n)$. The value of the splay tree as an access-sensitive search structure is that it can offer sequence time complexity approaching $O(n)$ if the access distribution of keys is highly nonuniform. By contrast, a standard balanced BST (e.g., AVL, red-black, etc.\ \cite{bst}) provides no access-distribution sensitivity and thus can be expected to require $O(n \log n)$ time to perform a sequence of $O(n)$ operations. 

A natural question is whether it is possible to combine the access-sensitive properties of the splay tree with the efficient worst-case properties of a balanced BST so that an $O(\log n)$ deadline time interval can be enforced while also minimizing the number of computations performed within that interval. In the next section we show that the {\em priority search tree} \cite{pst}, which was introduced in the same year as the splay tree (1985), can be repurposed in the form of an {\em access-adaptive} priority search tree (AAPST) to simultaneously exploit nonuniform key accesses while enforcing an $O(\log n)$ process-completion time bound.

\section{Adaptive Priority Search Tree}

The {\em priority search tree} (PST) is a data structure introduced by Edward McCreight in 1985 with the objective of storing a set of $n$ points in $\mathbb{R}^2$ in a way that allows for $O(\log n)$ update complexity, i.e., insertion or deletion of a point, and $O(\log n + k)$ complexity for semi-infinite 2-dimensional range queries where $k$ is the number of returned objects. This complexity is achieved by maintaining the points simultaneously in BST order on the $x$ coordinates and heap order on the $y$ coordinates within the same binary tree structure. The data structure allows for five main operations on a dataset $D$ of ordered pairs to be performed efficiently:
\begin{enumerate}
\item Insert an ordered pair into $D$.
\item Delete an ordered pair from $D$.
\item Given integers $x_0$, $x_1$, and $y_1$, among all the pairs $(x, y)$ in $D$ such that $x_0 \leq x \leq x_1$ and $y \leq y_1$, find the pair whose $x$ is minimal.
\item Given integers $x_0$ and $x_1$, among all the pairs $(x, y)$ in $D$ such that $x_0 \leq x \leq x_1$, find the pair whose $y$ is minimal.
\item Given integers $x_0$, $x_1$, and $y_1$, enumerate all $k$ pairs $(x, y)$ in $D$ such that $x_0 \leq x \leq x_1$ and $y \leq y_1$.
\end{enumerate}
The priority search tree was the first data structure to support 2-dimensional spatial search queries within the same $O(\log n + k)$ complexity of 1-dimensional range queries offered by balanced binary search trees (BSTs) while also supporting $O(\log n)$ update operations. Specifically operations 1-4 have time complexity $O(\log n)$ and operation 5 has $O(\log n + k)$ complexity. 

In his paper, Edward McCreight introduces two varieties of priority search trees, the first of which is the radix priority search tree. Each node in a radix PST contains exactly one ordered pair $(x, y)$. A radix PST is constructed so that the $y$-value of every child node is less than or equal to that of its parent in the case of a maximum PST, and greater than or equal to in the case of a minimum PST. In both cases, the $x$-value of every node in a right subtree is strictly less than that of every node in the left subtree. Furthermore, the cardinality of a node’s right subtree is equal to or one less than the node’s left subtree, ensuring balance. Unfortunately, this type of PST is not well suited for restructuring as a change to a singular node's $y$-value could require reconstruction of the entire tree in order to maintain its fidelity conditions. However, the second type of priority search tree, the balanced priority search tree, provides a solution to this issue. Unlike a radix PST node, a balanced PST node can contain up to two ordered pairs, the Q pair and the P pair. The Q pair $(Q_x, Q_y)$ is required for every node and it is selected for its near-median $x$-value, similar to a standard balanced binary tree. The P pair $(P_x, P_y)$ is optional and is chosen for its minimal $y$-value in the case of a minimum PST and maximal $y$-value in the case of a maximum PST, similar to a heap. Each pair $(x, y)$ in the set of keys appears as the Q pair of exactly one node $t$, and may also appear as the P pair of at most one ancestor node of $t$. This structure allows for more efficient updating of a node's $y$-value.

\bigskip
\noindent \begin{tabular}{ | c || c | c | }
\hline
\multicolumn{3}{| c |}{AOPST Node} \\
\hline
Field & Data Type & Description \\
\hline
Q x & Comparable & Key selected for near-median value of subtree\\
Q y & Integer & Priority associated with Q x \\
P x & Comparable & Key associated with maximal y-value in subtree \\
P y & Integer & Maximal priority in subtree \\
Valid P & Boolean & Whether the P-pair is valid \\
Left & AOPST Node & Left child \\
Right & AOPST Node & Right child \\
Parent & AOPST Node & Parent node \\
\hline
\end{tabular}
\bigskip

While the priority search tree was originally created to store two-dimensional coordinates, we introduce here an alternative use of the data structure. With some slight construction and operation alterations, the balanced priority search tree can be used as a distribution-sensitive search structure. We will call this specialized structure an {\em access adaptive priority search tree} (AAPST). The AAPST stores each key of a dataset in the $Q_x$-value of a node and its respective access frequency as the $Q_y$-value. Additionally, a key and its respective access frequency may be stored in the $P_x$ and $P_y$ fields of an ancestor node such that the non-null $P_y$ values form a max heap.

In other words, the skeleton of the tree maintains a balanced BST ordering of the keys while the access-frequencies associated with the keys are maintained in heap order. By slightly altering the search algorithm of a regular priority search tree, any search key can be found (or not found) in an AAPST in $O(\log n)$ time. The principal change to the standard balanced PST is the incrementing of the priorities (access frequencies) associated with the keys. Specifically, when a key is accessed by either an update or a query, its access count is incremented and the key's position in the heap may then be updated. This procedure is unique to the AAPST; it is not native to the standard priority search tree. As previously mentioned, a balanced PST allows for the insertion and deletion of an ordered pair from the tree, but these actions involve adding/removing both the P and Q pair from the tree. Since incrementing the priority of a key in an AAPST only affects the P pair's position in the heap portion of the tree, deleting and reinserting the Q pair is redundant and adds unnecessary comparisons. Thus, we designed a unique algorithm for repositioning only a key's P pair. This modified access/query algorithm can be defined as follows:

\begin{enumerate}
\item Find the query key and increment its associated priority.
\item If the query key was found in the P field of a node, and the incremented priority does not exceed the $P_y$ of its parent node, then return. If the incremented priority does exceed the $P_y$ of its parent node, then delete the P pair and reinsert it to the proper position.
\item If the query key was found in the Q field of a node, and that node's parent's P pair is invalid or has a lower priority than the found key, insert the Q pair as a P pair in the proper position.
\end{enumerate}
Step 1 takes time proportional to the pair's depth in the tree, and this will be the complexity of the operation in all cases 
in which the updated priority does not affect the heap order; otherwise, the complexity of the operation will be 
dominated by the $O(\log n)$ complexities of the P pair delete and insert algorithms. This establishes the worst-case $O(\log n)$ 
complexity of the new adaptive query algorithm. 

\begin{figure}[p]
\begin{center}
\begin{minipage}{0.45\textwidth}
\begin{figure}[H]
\begin{center}
\includegraphics[width=0.9\textwidth]{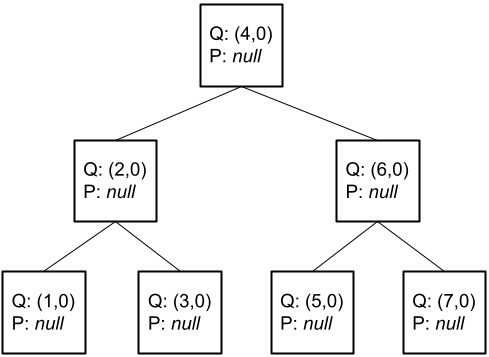}
\caption{\footnotesize A visual representation of an Access-Adaptive Priority Search Tree before any queries have been completed.}
\label{fig:tree}
\end{center}
\end{figure}
\end{minipage}\hfill
\begin{minipage}{0.45\textwidth}
\begin{figure}[H]
\begin{center}
\includegraphics[width=0.9\textwidth]{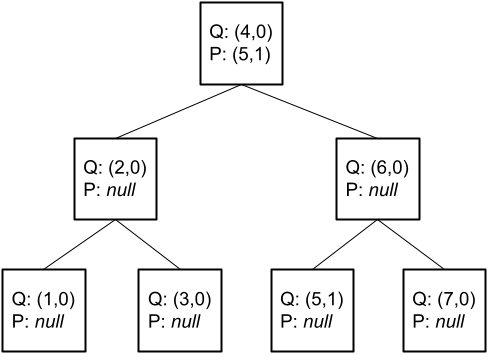}
\caption{\footnotesize The AAPST from \ref{fig:tree} after it is queried for the number 5. This query required 3 comparisons to find the key, for a total operation cost of 3 comparisons.}
\label{fig:query1}
\end{center}
\end{figure}
\end{minipage}
\end{center}
\end{figure}

\begin{figure}
\begin{center}
\begin{minipage}{0.4\textwidth}
\begin{figure}[H]
\begin{center}
\includegraphics[width=0.9\linewidth]{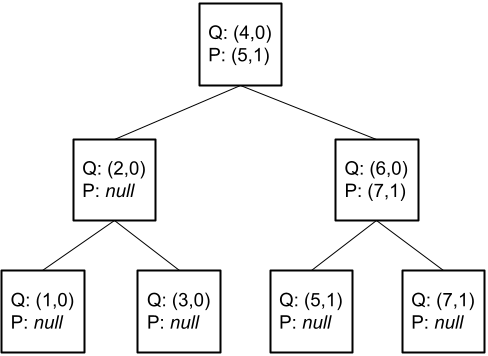}
\caption{\footnotesize  The AAPST from \ref{fig:query1} after it is queried for the number 7. This query required 4 comparisons to find the key and one comparison to insert the found key and its access count as a P pair, for a total operation cost of 5 comparisons.}
\label{fig:query2}
\end{center}
\end{figure}
\end{minipage}\hfill
\begin{minipage}{0.45\textwidth}
\begin{figure}[H]
\begin{center}
\includegraphics[width=0.9\textwidth]{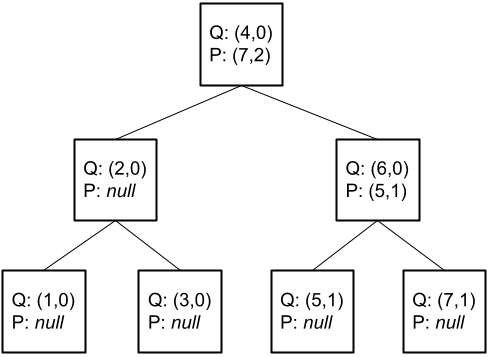}
\caption{\footnotesize The AAPST from \ref{fig:query2} after it is queried for the number 7. This query required 3 comparisons to find the key, one comparison to delete the p pair, and one comparison to reinsert the found key and its access count as a P pair, for a total operation cost of 5 comparisons.}
\label{fig:query3}
\end{center}
\end{figure}
\end{minipage}
\end{center}
\end{figure}

\begin{figure}
\begin{center}
\begin{minipage}{0.45\textwidth}
\begin{figure}[H]
\begin{center}
\includegraphics[width=0.9\textwidth]{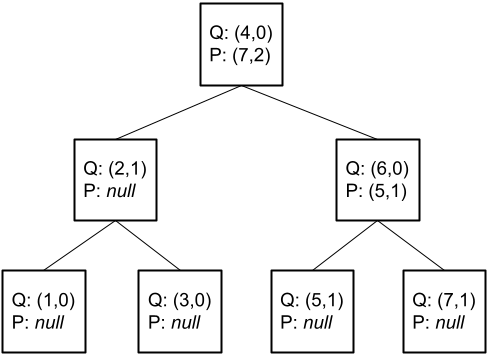}
\caption{\footnotesize The AAPST from \ref{fig:query3} after it is queried for the number 2. This query required 3 comparisons to find the key and one comparison to attempt to insert the found key and its access count as a P pair, for a total operation cost of 4 comparisons. }
\label{fig:query4}
\end{center}
\end{figure}
\end{minipage}\hfill
\begin{minipage}{0.45\textwidth}
\begin{figure}[H]
\begin{center}
\includegraphics[width=0.9\textwidth]{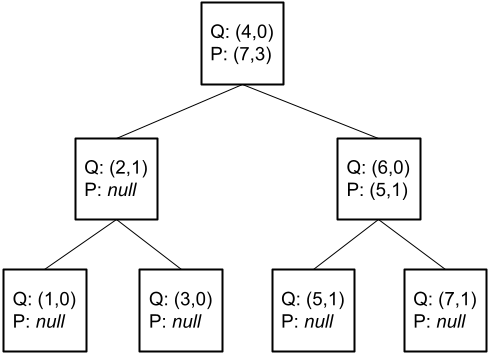}
\caption{\footnotesize The AAPST from \ref{fig:query4} after it is queried for the number 7. This query required 1 comparison to find the key, for a total operation cost of 1 comparison}
\label{fig:query5}
\end{center}
\end{figure}
\end{minipage}
\end{center}
\end{figure}

\begin{figure}
\begin{center}
\begin{minipage}{0.45\textwidth}
\begin{figure}[H]
\begin{center}
\includegraphics[width=0.9\textwidth]{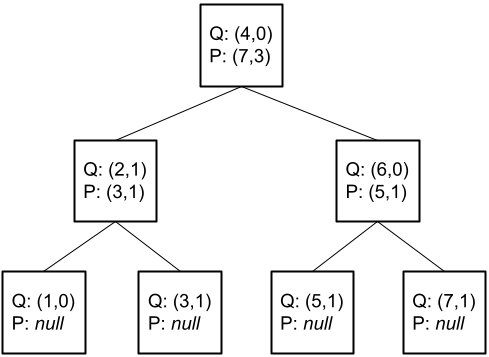}
\caption{\footnotesize The AAPST from \ref{fig:query5} after it is queried for the number 3. This query required 4 comparisons to find the key and one comparison to insert the found key and its access count as a P pair. }
\label{fig:query6}
\end{center}
\end{figure}
\end{minipage}\hfill
\begin{minipage}{0.45\textwidth}
\begin{figure}[H]
\begin{center}
\includegraphics[width=0.9\textwidth]{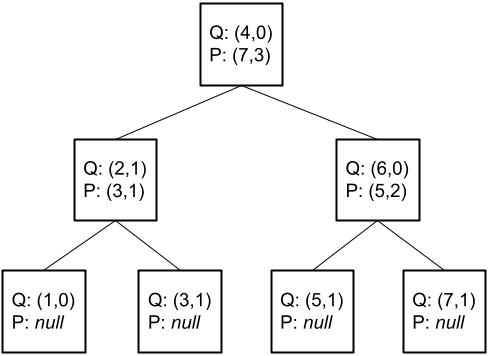}
\caption{\footnotesize The AAPST from \ref{fig:query6} after it is queried for the number 5. This query required 3 comparisons to find the key and one comparison to attempt to reinsert the found key and its access count as a P pair. }
\label{fig:query7}
\end{center}
\end{figure}
\end{minipage}
\end{center}
\end{figure}

Figures \ref{fig:tree} through \ref{fig:query7} visualize an AAPST and the effects this algorithm has on it as a set of keys is queried. The figure captions also note the number of comparisons required by each phase of the query process.

In the next section we provide practical comparisons of the AAPST and splay tree in the form of simulation results with varying degrees of non-uniformity in the sampling of query keys.

\section{Comparative Performance Results}

In this section we examine the relative performance characteristics of a conventional balanced binary search tree (BST), a splay tree, and the AAPST. In the case of uniformly sampled query keys, the splay tree and AAPST incur extra overhead compared to the BST. In the case of the splay tree, this overhead takes the form of extra comparisons performed as the tree is restructured. In the case of the AAPST, the overhead takes the form of extra comparisons to delete and insert the P pair as well as an extra key comparison per node visited: one comparison to the key in the $P_x$ position, and another comparison to the $Q_x$ key if it was not found in the P pair. Therefore, the goal of our tests is to examine how relative number of key comparisons used during the search of each data structure is affected by the distribution of queried keys. We should expect the BST to be superior in the uniform case while the splay tree and AAPST should perform better as the distribution becomes increasingly nonuniform. 

We define a value $p$, $0\leq p\leq 1$, to parameterize our key-access testing distributions with $p=0$ representing a uniform random distribution of key accesses; $p=1$ representing an exponentially-distributed sequence of accesses with the most frequently-accessed key representing approximately $50\%$ of the accesses, the next representing $25\%$ of the accesses, etc., such that $O(\log n)$ of the keys comprise $O(n)$ of the accesses; and $0<p<1$ representing a weighted mixture of accesses from the two distributions. 
	
\begin{figure}[H]
\begin{center}
\includegraphics[width=\figwid,keepaspectratio]{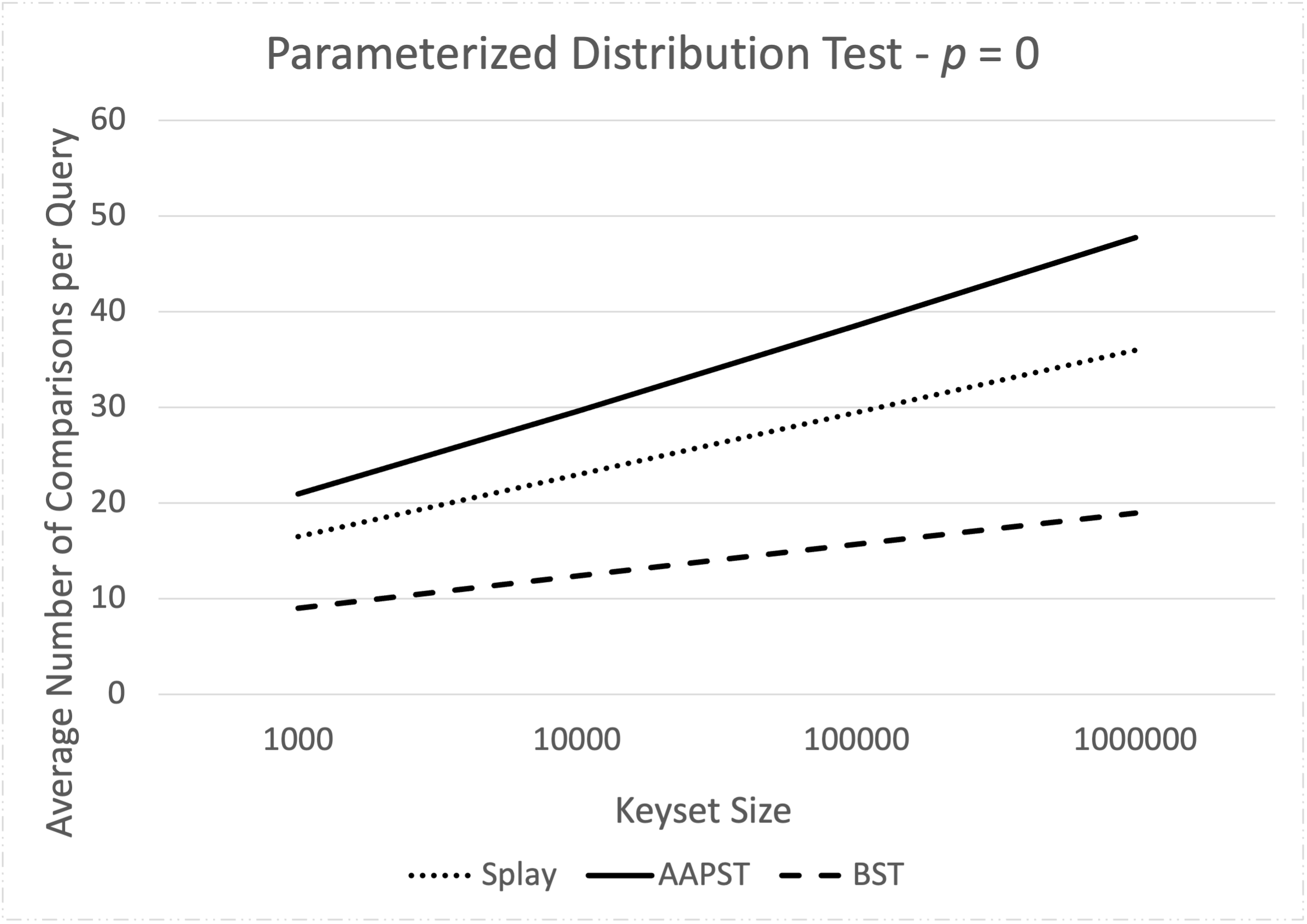}
\caption{\footnotesize This figure shows the average number of key comparisons for query keys drawn from uniform distribution for datasets of increasing size $n$. The expected number of key comparisons performed by the BST is $\log(n)$ while the splay and AAPS trees perform roughly twice as many comparisons per query.}
\label{fig:p-0}
\end{center}
\end{figure}

\begin{figure}[H]
\begin{center}
\includegraphics[width=\figwid,keepaspectratio]{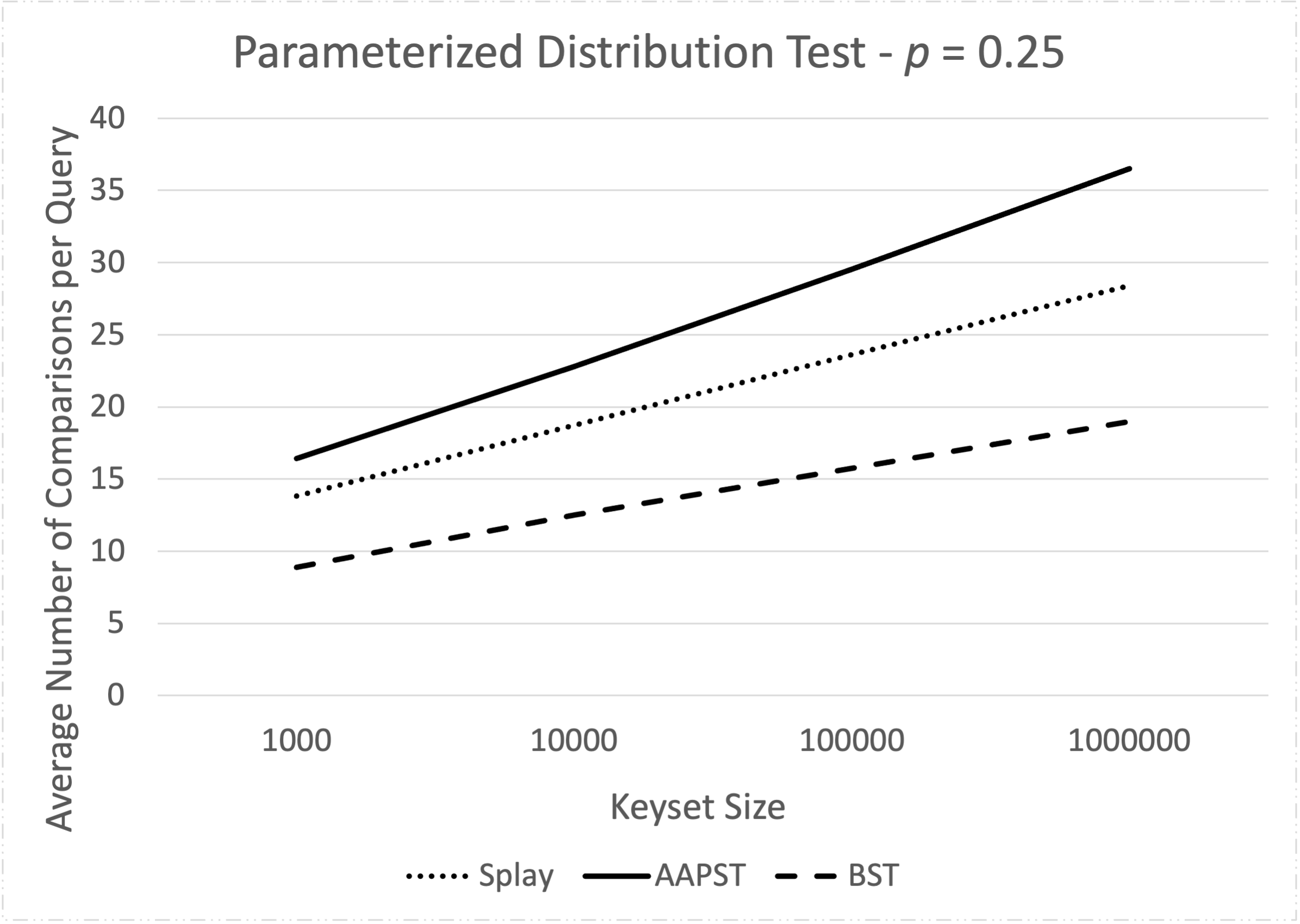}
\caption{\footnotesize This figure shows the average number of key comparisons when the $1/4$ of the query keys are sampled with exponential frequency and the remaining are sampled uniformly.}
\label{fig:p-25}
\end{center}
\end{figure}

\begin{figure}[H]
\begin{center}
\includegraphics[width=\figwid,keepaspectratio]{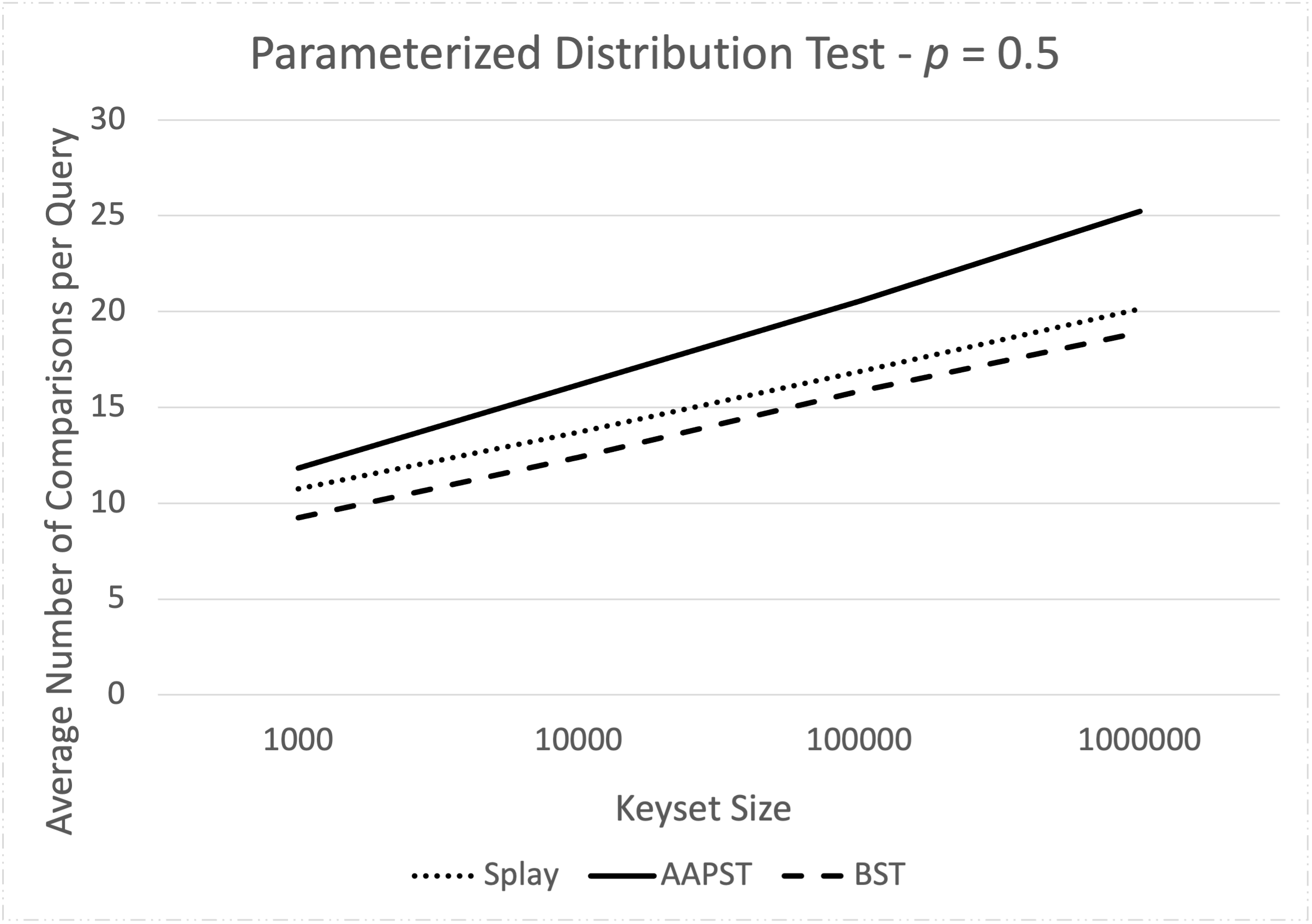}
\caption{\footnotesize This figure shows the average number of key comparisons when half of the query keys are sampled with exponential frequency and the remaining half are sampled uniformly.}
\label{fig:p-5}
\end{center}
\end{figure}

\begin{figure}[H]
\begin{center}
\includegraphics[width=\figwid,keepaspectratio]{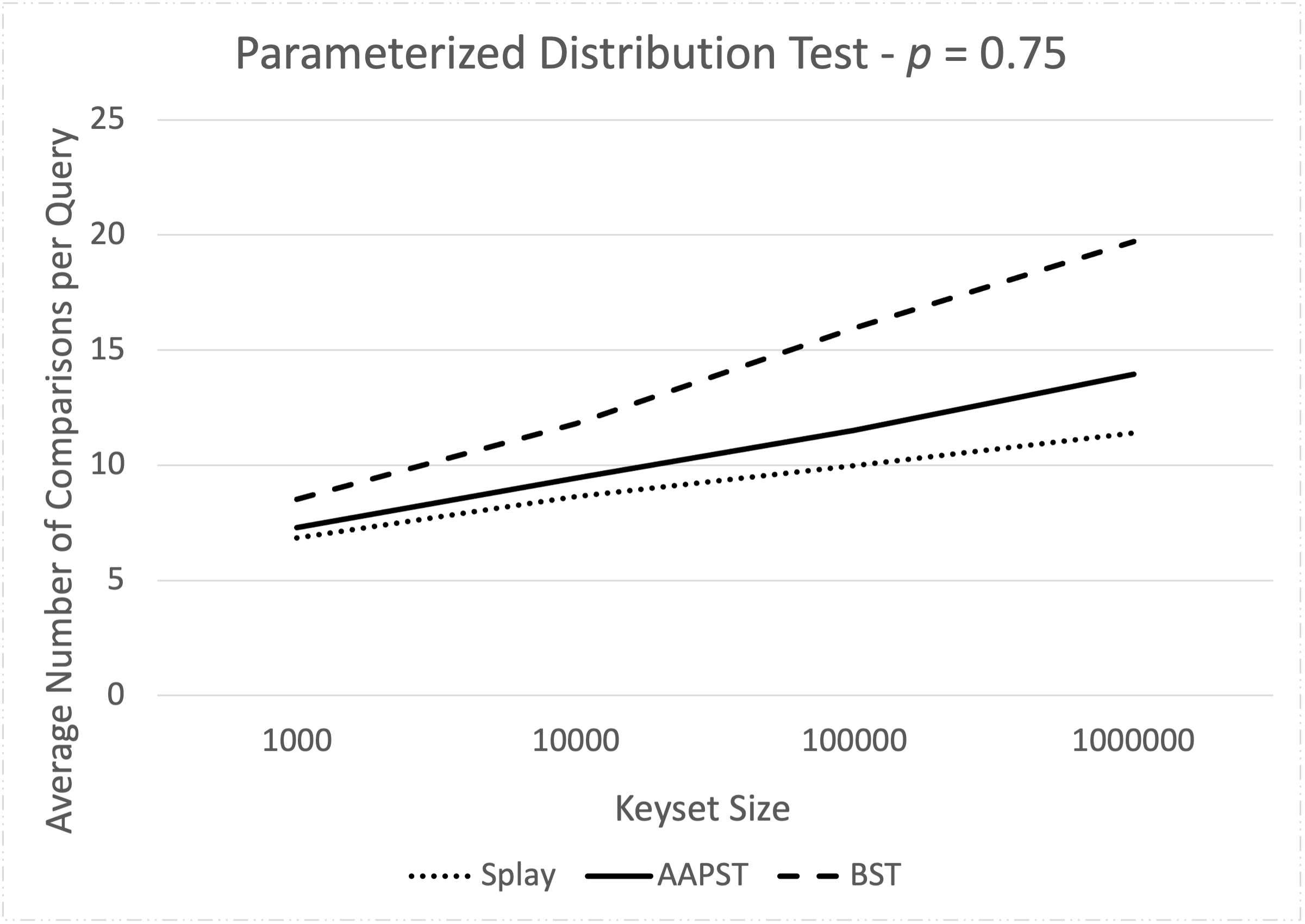}
\caption{\footnotesize This figure shows the average number of key comparisons when $75\%$ of the query keys are sampled according to an exponential distribution.}
\label{fig:p-75}
\end{center}
\end{figure}

\begin{figure}[H]
\begin{center}
\includegraphics[width=\figwid,keepaspectratio]{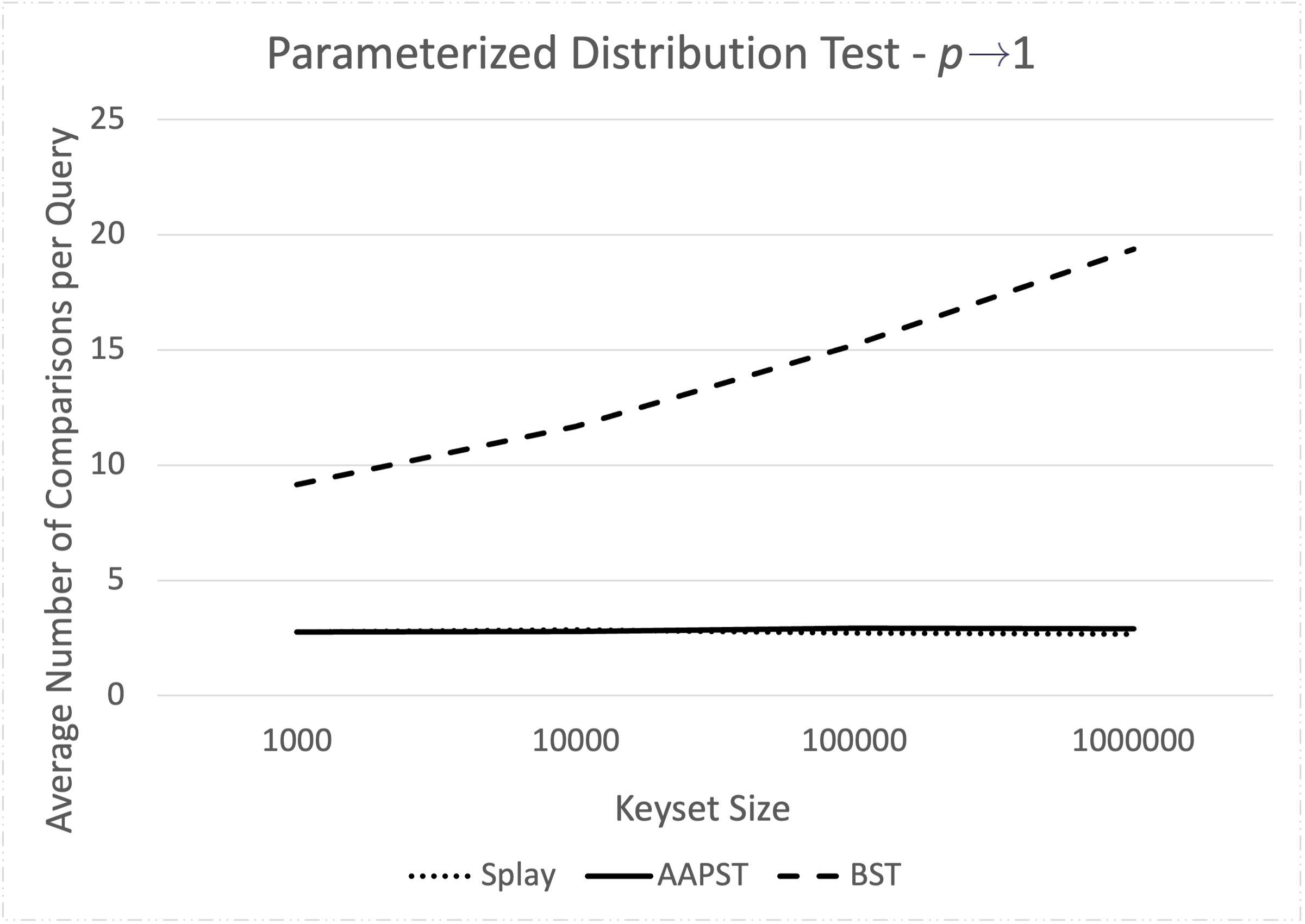}
\caption{\footnotesize This figure shows the average number of key comparisons when all query keys are sampled according to an exponential distribution, i.e., the frequency of access of different keys decreases exponentially.}
\label{fig:p-1}
\end{center}
\end{figure}

As can be seen in Figure \ref{fig:p-0}, the splay and AAPS trees perform comparably but are outperformed by the BST because of its lower overhead. In other words, the overhead of adaptivity incurred by the splay and AAPS trees does not yield dividends in the case of keys that are queried uniform-randomly. Figure \ref{fig:p-25} shows that the relative performance advantage of the BST decreases when $25\%$ of the key accesses exponentially distributed access frequencies. Figure \ref{fig:p-5} shows that when there is an equal mix of keys sampled from the uniform and exponential distributions the three search structures perform comparably. In the case of all keys sampled exponentially, Figures \ref{fig:p-75} and \ref{fig:p-1} show that the distribution-sensitivity properties of the splay and AAPS trees provide a significant performance advantage over the BST as the access frequencies tend toward an exponential distribution.

The test results presented in this section have focused on characterizing the relative performance of the three data structures in terms of a few key parameters. In the next section we present test results that are specific to applications involving a fixed process-completion time interval. 

\section{Test with Process-Completion Time Interval}

The results of the previous section corroborate what should be expected: adaptivity provides increasing efficiency benefits as the key-access distribution deviates increasingly from uniform-random sampling. At $p=0.75$, AAPST and Splay significantly outperform a static balanced BST, with this performance gap increasing between the adaptive trees and the BST as $p$ approaches $1$. In this section we focus on the principal use case for AAPST, which is to provide adaptive efficiency benefits with worst-case bounds for online and real-time applications. 

Given the scenario of the previous section for a keyset of size of $100k$, the worst-case number of comparisons per query of a balanced BST can be determined {\em a priori} to be $\lceil\log_2(100k)\rceil\,=\,17$. The worst-case for AAPST is $4\lceil\log_2(100k)\rceil-2\,=\,66$, which derives from the fact that $2\lceil\log_2(n)\rceil-1$ comparisons are needed to find a given key, and an additional $2\lceil\log_2(n)\rceil-1$ comparisons may be performed when the key is deleted and reinserted with its incremented priority. From a practical efficiency perspective, the complexity of this step is proportional to the {\em depth} of the key rather than the {\em height} of the tree, so repeated accesses to keys with high priorities near the root will not typically exhibit worst-case performance.

\begin{figure}[H]
\begin{center}
\includegraphics[width=\figwid,keepaspectratio]{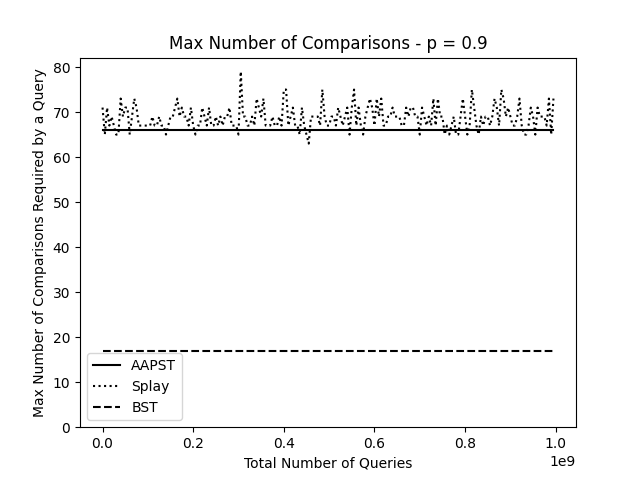}
\caption{\footnotesize This figure shows the maximum number of key comparisons performed in a single query in a set of 5 million queries where $90\%$ of the query keys are sampled according to an exponential distribution}
\label{fig:max}
\end{center}
\end{figure}

Figure \ref{fig:max} plots the maximum number of comparisons required during each of consecutive sequence of $5$ million queries. As should be expected, BST encounters a query during each interval that reaches its known upper bound number of key comparisons. The amortized complexity of Splay does not generally permit an $o(n)$ bound to be determined, so its worst-case query performance per sequence is highly variable and resembles that of a univariate normal distribution. This suggests that the maximum observed number of comparisons across all sequences would tend to increase with the number of sequences up to the trivial upper bound of $n$. The AAPST's worst-case of 66 comparisons is also not observed in this sequence of queries. This can be attributed to the fact that the tree's worst-case scenario only occurs in a highly specific situation that is unlikely during a set of uniform key accesses and becomes exceedingly rare as those accesses become more non-uniform. Regardless, figure \ref{fig:max} makes it clear that the AAPST's defined upper bound is still lower than many maximum access counts observed with the splay, and its observed upper bound is considerably lower. 

\begin{figure}[H]
\begin{center}
\includegraphics[width=\figwid,keepaspectratio]{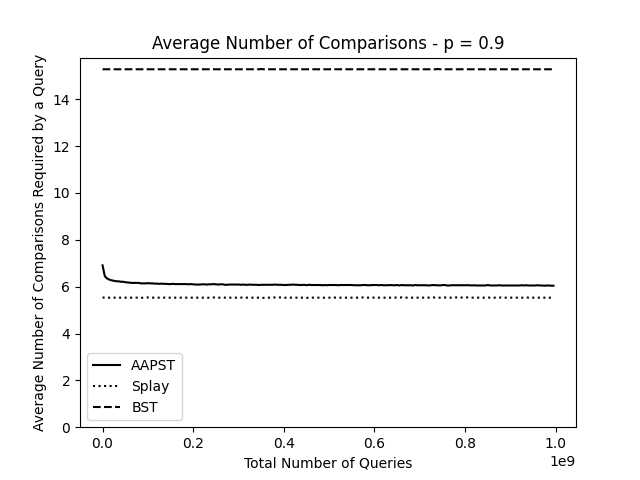}
\caption{\footnotesize This figure shows the average number of key comparisons per query in a set of 5 million queries where $90\%$ of the query keys are sampled according to an exponential distribution}
\label{fig:avg}
\end{center}
\end{figure}

Figure \ref{fig:avg} plots the average number of comparisons per sequence of $5$ million queries. The most notable feature of Figure \ref{fig:avg} is how close the averages are for AAPST and splay\footnote{The steady-state average number of comparisons for AAPST is very slightly higher than that of Splay. Future work will examine whether the separate find and priority-update (delete and reinsert) steps of the AAPST search algorithm can be integrated to potentially eliminate a redundant comparison per node.}. 

An interesting feature of both figure \ref{fig:max} and figure \ref{fig:avg} is that the comparison count for the AAPST is higher in the beginning of the query sequence and then drops down and flattens out into a consistent range as more queries are completed. This is due to the fact that the AAPST is initialized with all priorities equal to zero, which means that each key is at random depth in the tree. This leads to many more comparisons per query until frequently-accessed keys migrate to their steady-state depth near the top of the tree. In summary, the test scenario of Figure \ref{fig:max} provides a concrete example of how the AAPST can provide superior average-case performance bounds while simultaneously enforcing a strictly-defined worst-case optimal upper bound.

\section{Discussion}

It is not unusual for the standard metrics applied for assessing algorithms in the literature to fail to capture salient characteristics of interest in important, though in some cases specialized, practical applications. The amortized performance characteristics of the splay tree are certainly of significant theoretical importance, but its practical use is highly constrained by the fact that most query-retrieval applications involve interactive or related real-time constraints. In this paper we have emphasized that some important applications, such as in reliability-critical space systems, include both fixed-bound performance constraints and a need to minimize overall computational costs so as to reduce energy consumption and/or unwanted heat generation. With this in mind, we adapted the classical priority search tree to create a new access-adaptive priority search tree (AAPST) which satisfies these joint performance goals.

Another contribution of this paper is an empirical examination of the performance properties of the AAPST in terms of key variables relating to the bounding of worst-case behavior and expected-case computational expenditures as determined by a measure of nonuniformity of the distribution of key accesses. Our simulation results characterize the cases in which the AAPST can be expected to provide performance advantages over the splay tree and standard binary search trees.    

In summary, our simulation results show that the AAPST offers comparable access-sensitive performance to the splay tree while strictly bounding the complexity of each operation. This is a property that is required for interactive applications that must impose strict constraints on the worst-case response time of each operation, e.g., as arise in space systems and real-time tracking \cite{crc4}. Future work will examine finer-grain performance characteristics of the AAPST and their relevance to practical applications. More generally,
we hope that this paper will promote greater awareness of secondary algorithmic performance attributes to complement conventional computational complexity 
characterizations.

\bibliographystyle{plain}

\noindent{\includegraphics[width=4in,keepaspectratio]{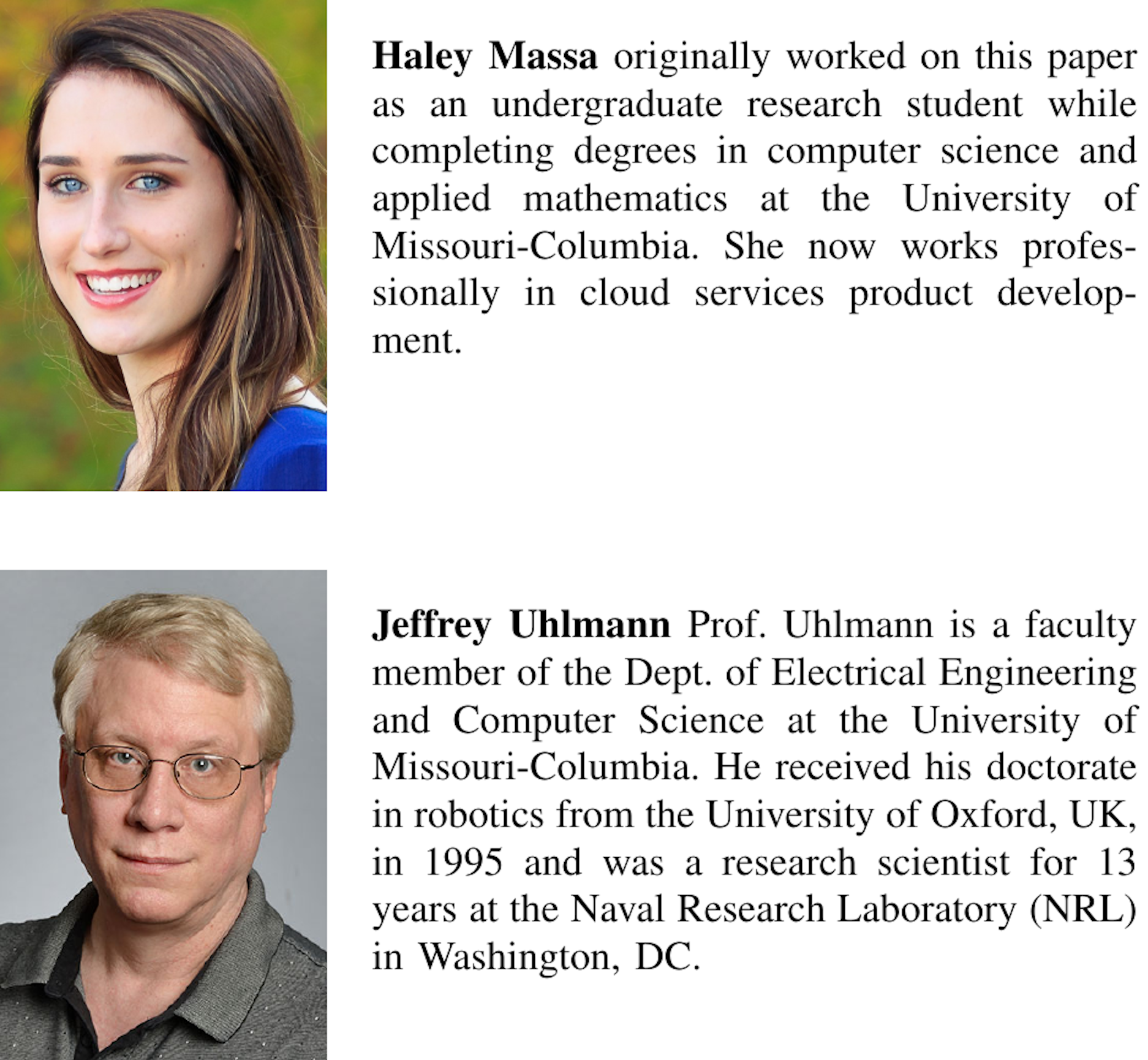}}

\end{document}